\documentclass[]{aa}  
\usepackage{soul}
\usepackage{graphicx}
\usepackage{txfonts}
\usepackage{xcolor}
\usepackage{subfig,graphicx}
\usepackage{ulem}
\usepackage[bookmarks=false]{hyperref}
\begin{document}
\nolinenumbers

\title{No Maunder Minimum phase in HD\,4915}

\author{
M. Flores-Trivigno\thanks{Based on observations made with ESO Telescopes at the La Silla Paranal Observatory under programmes ID 072.C-0488(E), 106.21DB.001, 105.20PH.001, 106.21TJ.001, 108.22CE.001, 109.2392.001 and 110.242T.001.
}\inst{1,2,3}
\and
A. P. Buccino\inst{3,4,8}
\and
E. González\inst{2,3}
\and
P. D.  Colombo\inst{3}
\and
C. González\inst{2}
\and
M. Jaque-Arancibia\inst{6,7}
\and
R. V. Ibañez Bustos\inst{5}
\and
C. Saffe\inst{1,2,3}
\and
P. Miquelarena\inst{1,2,3}
\and
J. Alacoria\inst{1,3}
\and
A. Collado\inst{1,2,3}}

\institute{Instituto de Ciencias Astron\'omicas, de la Tierra y del Espacio (ICATE), Espa\~na Sur 1512, CC 49, 5400 San Juan, Argentina\\ 
\email{matiasflorestrivigno@conicet.gov.ar}
\and
Facultad de Ciencias Exactas, F\'isicas y Naturales, Universidad Nacional de San Juan, San Juan, Argentina
\and
Consejo Nacional de Investigaciones Cient\'ificas y T\'ecnicas (CONICET), Argentina
\and
Instituto de Astronom\'ia y F\'isica del Espacio (IAFE), Buenos Aires, Argentina
\and
Laboratoire Lagrange, Université Côte d’Azur, Observatoire de la Côte d’Azur, CNRS, Boulevard de l’Observatoire, CS 34229, 06304 Nice Cedex 4, France
\and
Instituto de Investigación Multidisciplinar en Ciencia y Tecnología, Universidad de La Serena, Raúl Bitrán 1305,
La Serena, Chile
\and
Departamento de F\'isica y Astronom\'ia, Universidad de La Serena, Av. Cisternas 1200, La Serena, Chile
\and
Universidad de Buenos Aires, Facultad de Ciencias Exactas y Naturales, Departamento de Física. Buenos Aires, Argentina}

\nolinenumbers

   \date{Received September 15, 1996; accepted March 16, 1997}

 
  \abstract
{The long-term solar magnetic activity  and its cyclical behaviour, which is maintained by a dynamo mechanism, are both still challenging for the astrophysics. In particular, an atypical event occurred between 1645 and 1715 when the solar activity was remarkably decreased and the number of sunspots got extremely reduced. However, it is still unclear what happened to the solar cycle. The discovery of longer activity minima in cool stars may shed light on the nature of the complex mechanisms involved in the long-term behaviour of the solar-stellar dynamo.}
{Our aim is to explore if the G5V solar-like star HD\,4915, which showed a striking chromospheric activity pattern in a previous study performed with HIRES data, could be considered a bona fide Maunder Minimum (hereafter MM) candidate.}
{We have analyzed over 380 spectra acquired between 2003 and 2022 using HARPS and HIRES spectrographs. We carried out a detailed search  of activity signatures in HD\,4915 by using the Mount Wilson and the Balmer H$_{\alpha}$ activity indexes. This task was performed by means of the Generalised Lomb-Scargle periodogram.}
{The new HARPS data show that the chromospheric activity of HD 4915 is not decreasing. In fact, the rise of the activity after the broad minimum in three years gets to the level of activity before that phase, suggesting that it is not entering into a MM phase. We also calculate a rotation period of $23.4 \pm 0.2$ d, which has not yet been reported.}
   {HD\,4915 shows a distinctive activity behaviour initially attributed to a possible and incipient MM phase. The additional HARPS data 
   allow us to discard a MM in the star. Our analysis shows that the complex activity pattern of HD\,4915 could be ruled by a 
   multiple activity cycle, being a shorter cycle of 4.8-yr modulated by a potential longer one. More activity surveys with extensive records and suitable cadence are crucial for accurate identification of stars in Magnetic Grand Minima.}


   \maketitle
%

\nolinenumbers

\section{Introduction}

The stellar activity studies were initiated by Olin Wilson 
at the Mount Wilson Observatory. This pioneering investigation \citep{1978ApJ...226..379W} and subsequent works \citep{1978PASP...90..267V,1984ApJ...279..763N,Duncan91,1995ApJ...441..436G,1995ApJ...438..269B} showed that the chromospheric variations, similar to those observed in the Sun, are also common in other main sequence stars. Another important contribution to this field was made by \citet{1998ASPC..154..153B}. In their study, the authors examined over one thousand stars by measuring the Ca {\sc ii} H\&K line-core fluxes. As a result, they were able to identify some differences in the long-term activity behaviour: i) a group of the stars showed a cyclic behaviour, having periods between 2.5 and 25 yr with intermediate activity levels,
ii) some stars displayed an erratic behaviour, in general they were associated to stars with high activity levels, iii) the last group presented flat activity levels, corresponding to inactive stars. This group initially drew a lot of attention due to they are ideal targets to identify and to study stars in magnetic minima, such as 
the solar Maunder Minimum \citep{1976Sci...192.1189E}.

The records of sunspots from 1645 to 1715 revealed a solar surface    
with very few sunspots. During this period, widely known as Maunder
Minimum (hereafter MM), the solar activity was strongly weakened. 
Some authors suggest that the 11-year cycle was disrupted. In contrast, different works based on counts of sunspots and cosmogenic isotopes indicate a weaker but persistent cycle. Furthermore, 
besides of the MM phase, other prolonged intervals of low solar activity such as Dalton and Spöerer (hereafter DM and SM) have been identified, even using other alternative proxies like auroral and cosmogenic $^{14}$C data \citep[see][for more details]{2014JGRA..119.2379M,2015A&A...581A..95U,2021ApJ...909...29H}. These periods of pronounced changes of the long-term cycling activity are called Magnetic Grand Minima \citep[MGM,][]{2012IAUS..286..335S}. However, some differences in the magnetic topology 
seem to be present. For instance, it has been reported a strong magnetic hemispherical asymmetry
during the MM phase, while the DM phase would resemble a modern minima \citep{Vaquero15AdSR,Hayakawa21mnras,2021ApJ...909...29H}.

The search for stars in MGM states is very important for solar and stellar activity studies, they would allow us to improve the current theoretical solar-dynamo models \citep{2010LRSP....7....3C,2021A&A...645L...6F}.  In fact, although for the solar case the MGM are linked to a special state of the solar dynamo \citep{2015A&A...581A..95U,2021ApJ...909...29H}, at present, the origin of these unusual periods is still under discussion. In addition, it is not clear whether the solar MGM is a regular or chaotic process \citep{2017LRSP...14....3U,2023SSRv..219...19B}. These stars are also of considerable interest for studies on space weather and Earth's climate
\citep[e.g.,][]{2017JSWSC...7A..33O,2023AGUA....400964Y}.

Although some stars showing broad minimum phases in their activity registry are known \citep[e.g.,][]{Metcalfe13,2020A&A...640A..46F,Baum_2022}, currently only four stars have been reported as MM candidates;
HD\,217014 \citep{2009A&A...508.1417P}, HD\,4915 \citep{Shah_2018}, HD\,20807 \citep{2021A&A...645L...6F}, and HD\,166620 \citep{2022ApJ...936L..23L}. In particular, 
the time series of the Mount Wilson S-Index ($\mathrm{S}_\mathrm{{MW}}$) for HD\,4915 showed a clear decreasing amplitude of its magnetic activity cycle. As a result, 
\citet{Shah_2018} suggest that HD\,4915 is possibly entering in a MM state. In addition, they also note that the activity behaviour of the star show a strong similarity when compared with the transition of the Sun into the DM. Then, based on this evidence, the authors suggest continuing to observe the object to confirm the nature of this incipient MGM state.

HD\,4915  ($=$ HIP\,3979, V $=$ 6.9, B$-$V $=$ 0.67) is
a solar type G5V star with no planet detected at the moment 
\citep{2017AJ....153..208B}. This object has been extensively observed 
under the California Planet Search (CPS) Radial Velocity Survey \citep{2010ApJ...721.1467H} using the Keck-HIRES spectrograph. Fortunately, at the southern hemisphere, 
the extensive database of HARPS also provides a suitable amount of high-resolution spectra, which represents a unique opportunity to confirm or reject a MGM state (e.g., a MM or a DM) in HD\,4915 by means of two of the most used
activity proxies ($\mathrm{S}_\mathrm{{MW}}$ and H$_{\alpha}$). The importance of studying stars near their MM state strongly encourages us to perform an in-depth analysis of this remarkable object.

This work is organised as follows: In \S 2, the observations and data reduction are described. In \S 3  we describe our
main results. In \S 4  we outline our discussion, while our main conclusions are provided in \S 5.


\section{Observations and Data Reduction}
The stellar spectra for HD\,4915 were obtained from the European Southern Observatory (ESO) archive\footnote{\url{http://archive.eso.org/wdb/wdb/adp/phase3_spectral/form?phase3_collection=HARPS}}. These data were acquired with the HARPS spectrograph\footnote{\url{http://www.eso.org/sci/facilities/lasilla/instruments/harps/overview.html}}, installed at La Silla 3.6 m (ESO) telescope in Chile \citep{2003Msngr.114...20M}. The resolution of this fibre-fed spectrograph is $R$ $\sim$115\,000 with a spectral coverage ranging from 3782 \AA \, to 6913 \AA. We have collected more than 300 spectra with a resulting signal-to-noise (S/N) $\sim$150, measured near 6010 \AA. These observations were acquired between the years 2003 and 2022 under the ID programmes: 072.C-0488(E), 106.21DB.001, 105.20PH.001, 106.21TJ.001, 108.22CE.001, 109.2392.001 and 110.242T.001. They were processed with the DRS HARPS pipeline\footnote{\url{http://www.eso.org/sci/facilities/lasilla/instruments/harps/doc.html}}. 

Previous to the calculation of the activity indexes, the spectra were corrected for radial velocities by using the standard IRAF\footnote{IRAF is distributed by the National Optical Astronomical Observatories, which is operated by the Association of Universities for Research
in Astronomy, Inc. (AURA), under a cooperative agreement with the
National Science Foundation.} routines. Those spectra with  
low S/N (i.e., $\leq$100) were discarded. Then, the $S$-indexes were obtained following the methodology described in \citet{1978PASP...90..267V}. In summary, we first integrated the flux in two windows centred at the cores of the Ca {\sc ii} H\&K lines (3968.47 \AA \ and \ 3933.66 \AA, respectively), weighted with triangular profiles of 1.09 \AA \ full width at half-maximum (FWHM), and computed the ratio of these fluxes to the mean continuum flux, which was integrated in two passbands of $\sim$20 \AA \ width centred at 3891 and 4001 \AA \ \citep[see][for details]{2017MNRAS.464.4299F}. Then, we used the calibration of \citet{2011arXiv1107.5325L} to derive the $\mathrm{S}_\mathrm{MW}$ indexes. On the other hand, the calculation of the H$_{\alpha}$ indexes for HARPS spectra
was performed by using the code \textsc{ACTIN 2}\footnote{\url{https://github.com/gomesdasilva/ACTIN2}}, which has been implemented in a python code by \citet{2021A&A...646A..77G}.

\section{Results}
\subsection{Long-term activity}
\begin{figure}

	\includegraphics[width=0.4\textwidth]{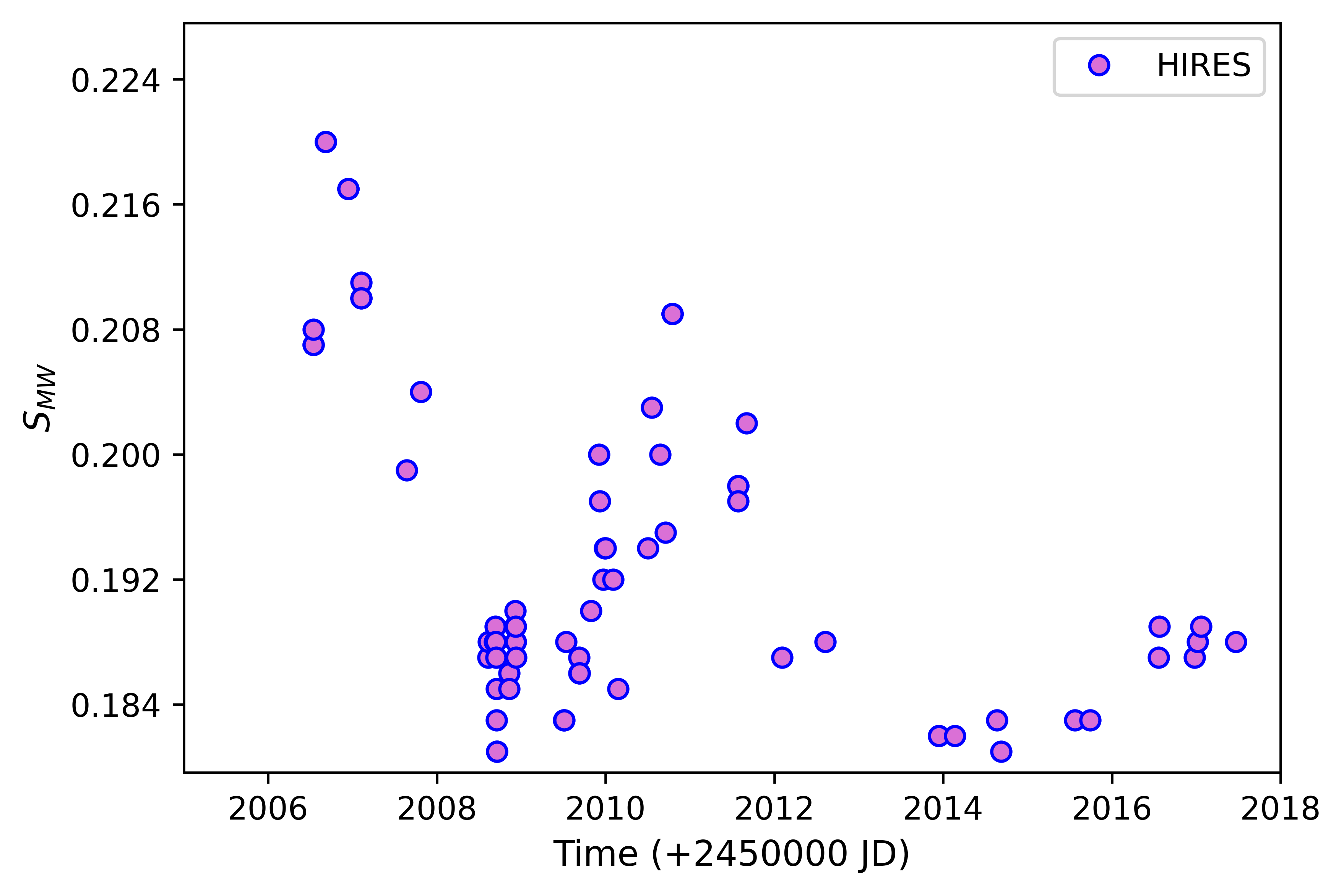}

	\includegraphics[width=0.4\textwidth]{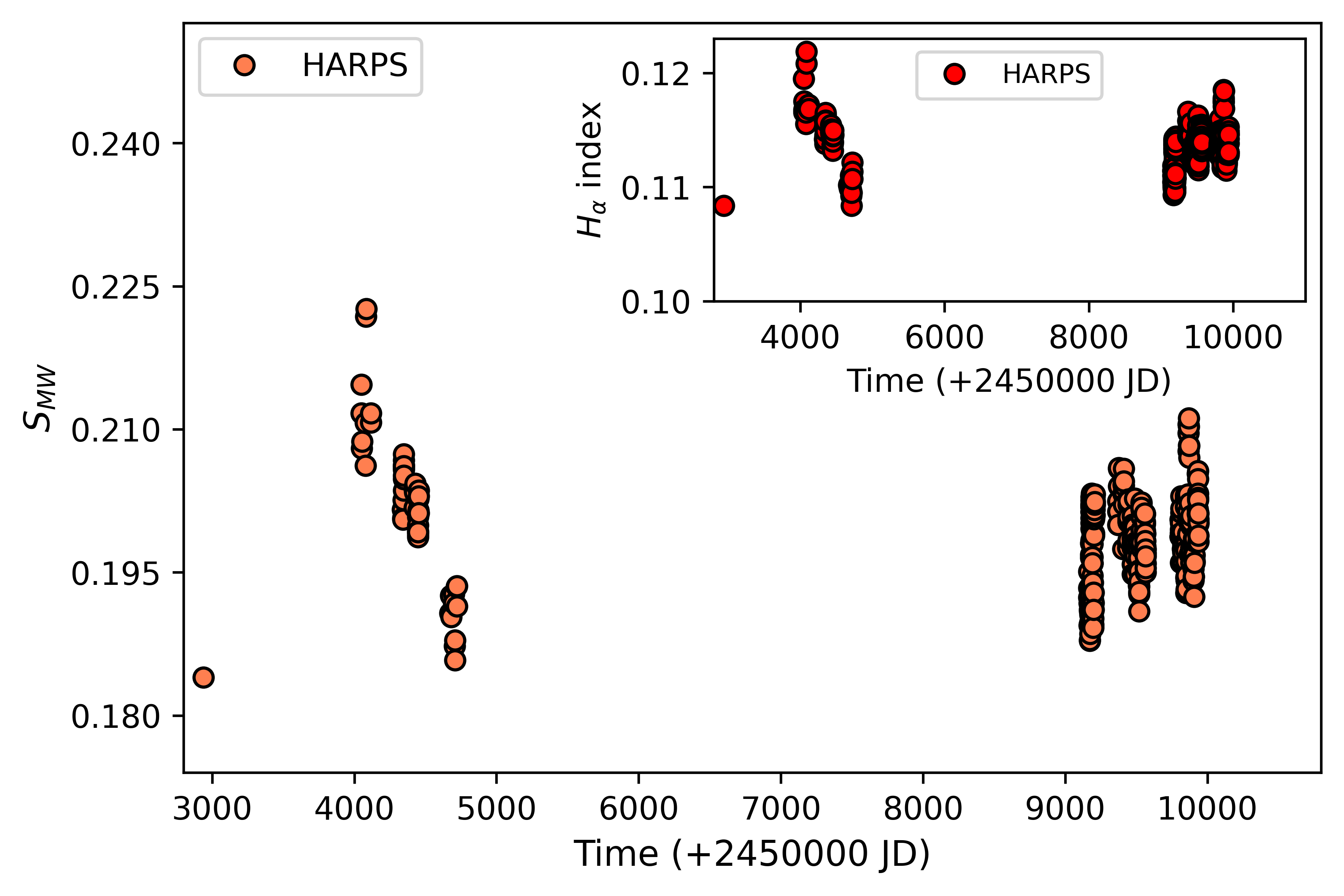}
	\includegraphics[width=0.4\textwidth]{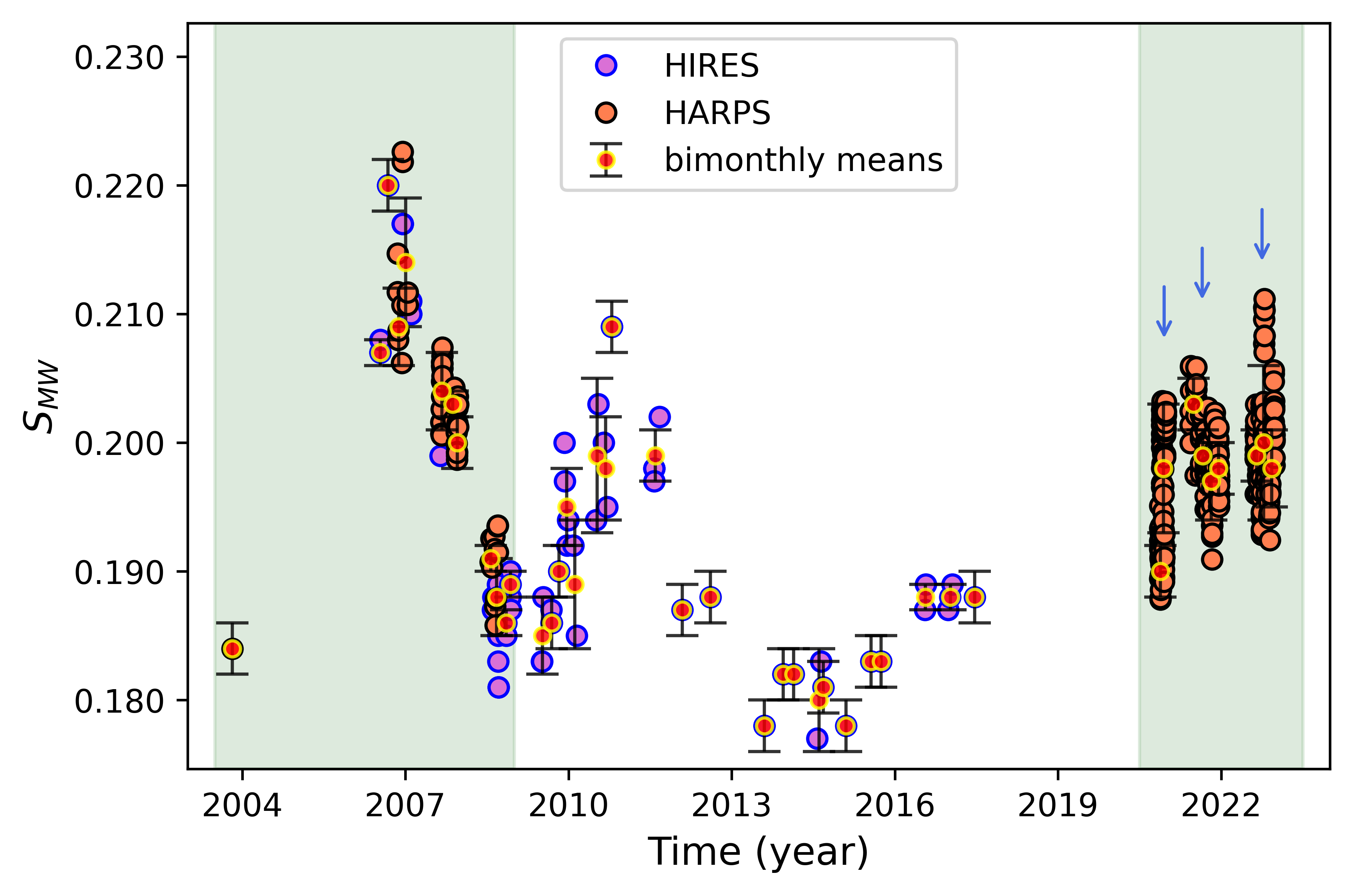}
	
\caption{$\mathrm{S}_\mathrm{{MW}}$ index variation of HD\,4915. Upper panel: Time series of the Mt. Wilson index
from HIRES observations. Middle panel: Similar to upper panel, but for HARPS data. This panel also includes the H$_{\alpha}$ indexes. Lower panel: Time series of the Mt Wilson index for HD\,4915 from combined observations. The HARPS bimonthly means are also indicated with red dots. Both shaded regions represent the time coverage of the new data acquired with HARPS. The blue arrows represent HARPS observations with the highest cadence (refer to Section 3.2 for details).
}
\label{Figone}
\end{figure}

In Fig.\ref{Figone} (upper panel) we show the time series of the $\mathrm{S}_\mathrm{{MW}}$ indexes derived from HIRES spectra, which have been taken from \citet{Shah_2018}. In this case, HD\,4915 displays a significant and striking decrease in the amplitude of the magnetic activity cycle, as pointed out by the authors. Then, in order to identify a possible MGM state in this star, in the middle panel we show the time series of the $\mathrm{S}_\mathrm{{MW}}$ indexes calculated from HARPS spectra. In addition, the inset plot corresponds to the H$_{\alpha}$ indexes that were calculated with \textsc{ACTIN 2} for these same data. We evaluate the strength of a possible correlation between these indexes by using the Bayesian framework implemented in Python code by \citet{2016OLEB...46..385F}. This code allows us to calculate the posterior probability distribution of the correlation coefficient $\rho$. As a result, we obtained a very high $\rho$ of 0.912 $\pm$ 0.01 with a 95\% credible interval between 0.894 and 0.931. Finally, in the lower panel we show the observations corresponding to both surveys, that is, HIRES and HARPS. We have also included the monthly average values of the combined indexes. It can be noted that the last HARPS data (i.e., the right shaded region) show a clear increase in the activity of HD\,4915, when compared to the third maximum (with the lowest amplitude) detected in the HIRES observations. 

\begin{figure}
 $\begin{array}{c}
 \includegraphics[scale=0.5]{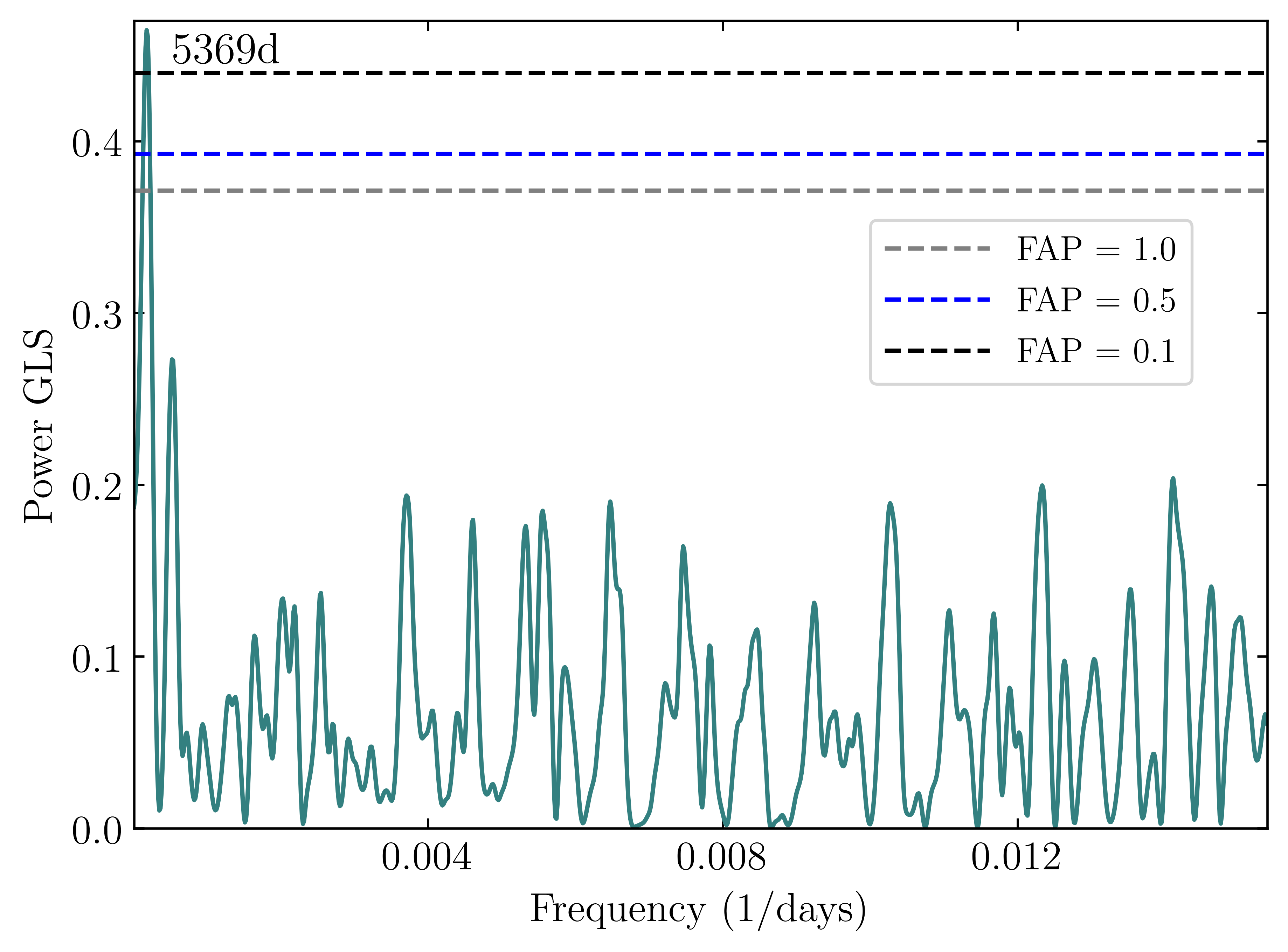} \\
 \includegraphics[scale=0.5]{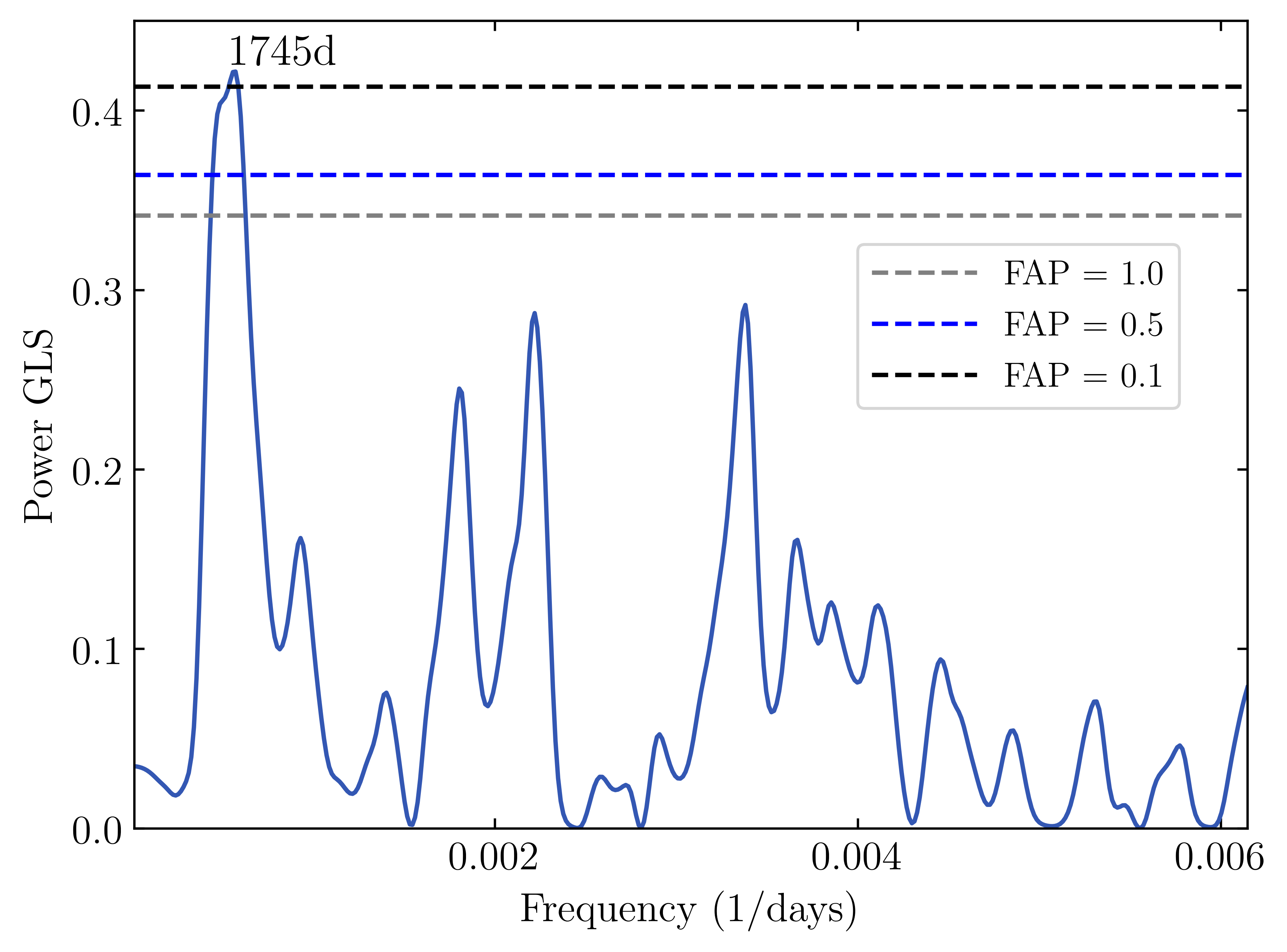}
 \end{array}$
 \centering
 \caption{Upper panel: GLS periodogram for the combined (bimonthly averages) data. Lower panel: GLS periodogram after subtracting the $\sim$14.7 yr long-term period. Different values of FAP are indicated with dashed horizontal lines.}
\label{fiot}
\end{figure}

\begin{figure}
    \includegraphics[scale=0.35]{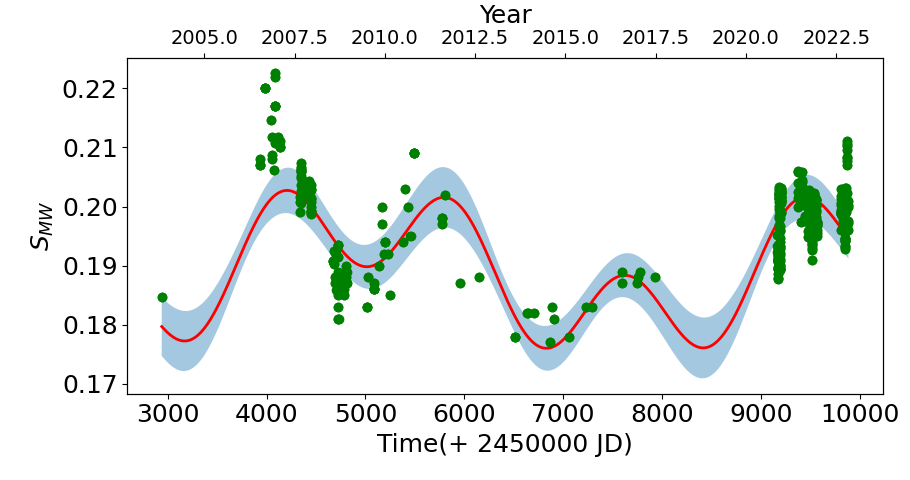}
    \caption{$\mathrm{S}_\mathrm{MW}$  derived from the HIRES and  HARPS spectra. The solid line is the least-square fit with a harmonic function of periods 1745 $\pm$ 114 days   and  5369 $\pm$ 787 days, derived from the GLS periodogram in Fig. \ref{fiot}, with a Pearson correlation coefficient of 0.65  and a p-value$\leq 10^{-4}$. The shaded region represents $\pm$3$\sigma$ deviations.}
    \label{doble}
\end{figure}

Based on the long-term chromospheric activity pattern showed by the HD\,4915 (see Fig.\ref{Figone}), we performed a period analysis for the combined data set. To do so, we applied the generalised Lomb-Scargle (hereafter GLS) periodogram and also calculate the false-alarm probability (hereafter FAP) of the periods, following \citet{2009A&A...496..577Z}. These tasks were carry out by using a python code, which implements the GLS\footnote{\url{https://github.com/mzechmeister/GLS/blob/master/python/gls.py}}. For this calculation we have used the bimonthly averages of all combined data. This procedure, also applied in previous works \citep[e.g.,][]{2010ApJ...723L.213M,2021A&A...645L...6F}, allows us to discard short-timescale variations associated to rotational modulation. As a result, we obtain a period of $P_1$=(5369 $\pm$787) d with an FAP of 4.0 x 10$^{-04}$. Moreover, in order to look for additional short-term yearly variations in the data, we subtracted the main periodic variation of 5369 d and recalculated the corresponding
GLS periodogram, according to \citet{2016A&A...595A..12S}. For this case, the period and the corresponding FAP are $P_2$=(1745$\pm$114) d and 7.5 x 10$^{-04}$, respectively. As in previous works \citep{2016A&A...589A.135F,SuarezMascareno16,2018MNRAS.476.2751F}, here we have adopted a
cut-off in FAP of 0.1 per cent (0.001) for a reliable periodicity. Upper and lower panels of Fig.\ref{fiot} correspond to the resulting GLS periodograms.  In Fig. \ref{doble} we show the $\mathrm{S}_\mathrm{{MW}}$ time series and the best-fit with a harmonic function of periods $P_1$  and $P_2$, with a Pearson correlation coefficient of 0.65 and a p-value$\leq 10^{-4}$.

\subsection{Rotation}
\begin{figure}
    \includegraphics[width=0.5\textwidth]{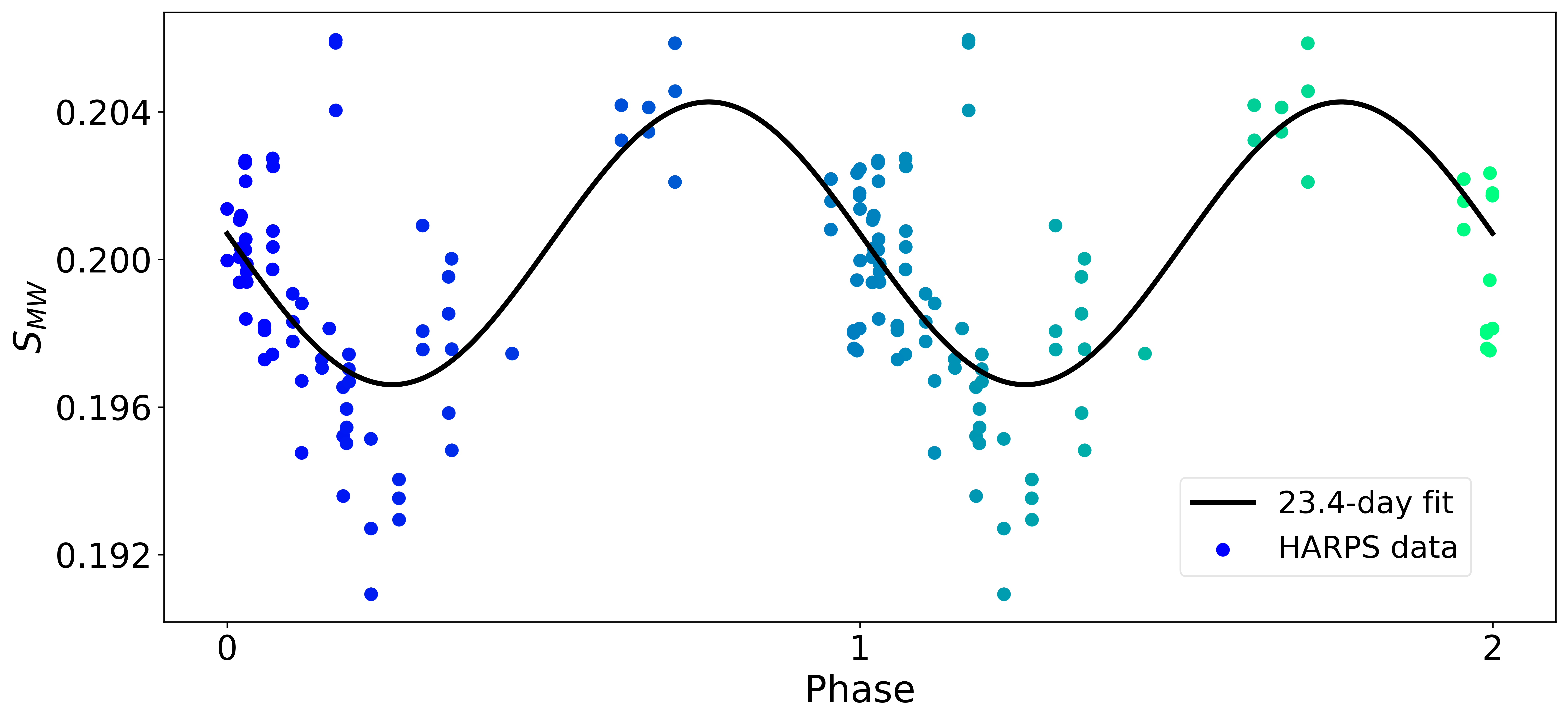}

\caption{Mount Wilson indexes from the second season highlighted in Fig. \ref{Figone}, phased (shown between 0 and 2) with a period of 23.4 d. The solid black line represents the 
 fitted harmonic curve.}

      \label{phased}
\end{figure}

The stellar rotation rate is one of the main dynamic parameters to characterise stellar activity (e.g. \citealt{Noyes84,Bohm07,Astudillo17}). To detect a possible rotational signature in HD\,4915, we isolated, from Fig.\ref{Figone}, the high-cadence Mount Wilson indexes derived from HARPS spectra (i.e., those corresponding to observations made from 2020 to 2022). In this way, we focused on the 
last three observing seasons: the first one corresponds to the interval 2020.9–2021.0, the second corresponds to the interval 2021.4–2021.9, while the last interval corresponds to 2022.6-2022.9 (see the blue arrows marked in the Fig. \ref{Figone}). Then, we computed the GLS periodogram for each of these seasons, where we detected a significant peak at $22.8 \pm 0.44$ d, $23.0 \pm 0.20$ d, and $24.3 \pm 0.27$ d respectively. As a result, we obtained a mean short-term period of $23.4 \pm 0.2$ d. In Fig. \ref{phased}, we present the observations for the second season adjusted according to our derived mean period. 
Although we show only the season with the rotation period closest to the mean and the lowest error (i.e., $\pm 0.20$ d), the other two seasons also demonstrate good agreement. Then, considering the level of activity of HD\,4915 ($log(R'_{HK})=-4.84$) and  the empirical relations in \cite{Noyes84} and \cite{Mamajek08}, we estimated a rotation period of  $\sim23.3 \pm 4.30$ d and $\sim23.2 \pm 3.32$ d respectively. As can be noted, our result agrees with the rotational periods derived from the chromospheric-activity calibrations.

\subsection{Ca {\sc ii} H\&K line-core fluxes during minimum phase}
Similar to \citet{2021A&A...645L...6F}, we  explored the components of the Ca {\sc ii} H\&K line-core fluxes responsible for the low $\mathrm{S}_\mathrm{{MW}}$ index of HD\,4915, in order to disentangle whether the chromospheric heating during the minimum phase 
is purely acoustic or magnetic. To do so, first we computed the Ca {\sc ii} surface  flux from the minimum $\mathrm{S}_\mathrm{{MW}}$ value using the relations from \cite{Mittag13} (Eq. 7-10). During the minimum phase, the surface flux gets to $F_{HK}=2.34\times 10^6$ erg s$^{-1}$ cm$^{-2}$. Following \cite{Martin17}, where the minimum  Ca {\sc ii}  flux is composed of the basal and the photospheric contributions, the minimum flux in the Ca {\sc ii} lines should be $1.61\times 10^6$ erg s$^{-1}$ cm$^{-2}$ for HD\,4915 ($B-V=0.671$). Thus during the minimum phase,  the excess flux in the Ca  {\sc ii} emission  of $7.3\times 10^5$ erg s$^{-1}$ cm$^{-2}$ could be attributed to purely magnetic activity.

\section{Discussion}

It is well known that the Sun has experienced MGM, such as the MM, the DM and the SM \citep{2017A&A...599A..58V,Shah_2018,2021MNRAS.504.5199C,2021JSWSC..11...17S}. Similar activity patterns have been also detected in four stars (HD\,217014, HD\,4915, HD\,20807, and HD\,166620). In particular, HD\,4915 was  extensively observed by HIRES between 2006 and 2018. These observations allow  \cite{Shah_2018} to detected a  minimum activity phase in HD\,4915 starting at 2013, which led them to suggest that the star is possibly entering in a MGM. In this work, the authors also emphasize the need to include additional data to confirm or reject the presence of this phenomenon in HD\,4915.

Taken into account the data for HD\,4915 available in the HARPS database, 
in the present work we extended the Mount Wilson time series of HD\,4915 to a time span of 19 years (2003-2022). The Mount Wilson indexes derived from HARPS spectra between 2005 and 2017 presented in Fig. \ref{Figone}, which are clearly replicated by line profile variations of  H$_{\alpha}$ (middle panel), coincide within the statistical errors with the indexes obtained from HIRES spectra reported in \cite{Shah_2018}, showing a correspondence between both datasets. Surprisingly, a direct inspection of Fig.\ref{Figone} (lower panel) shows a clear increase in the amplitude of the magnetic activity for HARPS data, after the last decrease of activity revealed by the HIRES observations. This fact rules out the possibility that the star HD\,4915 is entering in a MGM. This new time series confirmed that the broad minimum phase is abandoned near 2016 and in October 2020 the monthly level of activity rises  to $\langle\mathrm{S}_\mathrm{MW}\rangle=0.195$ similar to June 2011 before the minimum phase. Thus, this  broad minimum phase seems to last only for three years (2013-2016).  

To explore its long-term activity behaviour, we computed the GLS-periodogram for the  bimonthly means. Similar to other solar-type stars \citep{Bohm07,Metcalfe13,2016A&A...590A.133O,Mittag23}, HD\,4915 seems to present two co-existing activity cycles of $\sim$ 14.7 and $\sim$ 4.8 years. One way to test the confidence of both activity periods 
 is analyzing their location in the empirical diagrams rotation-activity available in the literature.  Given the rotation period of $\sim$23 days derived from the short-term modulation in HARPS index, the shorter activity cycle fits the inactive branch in \citeauthor{Bohm07}'s diagram, while the longer one is 60\% lower than the expected one for the active branch. However,  \cite{Boro18} expand the statistics on cyclic stars and put in doubt the active branch of \citeauthor{Bohm07}'s diagram. Recently, \cite{Mittag23} revisited the empirical relation between the stellar rotation period and the activity cycle lengths. The short activity cycle detected for HD\,4915 fit the short period in the power-law $P_{cyc}-P_{rot}$ relation corrected with colour $B-V$, where the expected $P_{cyc}^S\sim  $ 1945 days ($\sim$ 5.3 yr). The coincidence within the statistical error persists even under the colour correction where the expected value is $P_{cyc}^S\sim$ 1668  days. While the calculated long period  is $P_{cyc}^L\sim$  9100 days  with a 40 $\%$ dispersion, the detected long period of (5369 $\pm$ 787) days slightly overlaps with the calculated range.  Thus, although the long period is significant, we recommend to add new observations to the present time series to get a longer time-span and a more  accurate value. Nevertheless, our results suggest that the low activity reported between 2013 and 2016 could be associated to the minimum of a longer activity cycle which modulates the short prominent one.

In order to explore the level of magnetic contribution during the minimum phase,  we computed the surface calcium flux during this lapse. This value exceeds the basal and photospheric flux, which reinforces the fact that a purely magnetic component is responsible for the chromospheric heating in the broad minimum and the magnetic activity although depressed is still present, similar to the Dalton Minimum phase in the Sun.

\section{Conclusions}

\cite{Shah_2018} reported HD\,4915 as one of the four MM  solar-type candidates. However, they suggest that further observations of this star are necessary for conclusive results. In this sense, in the present work we included new public observations and perform a long-term analysis of the Mount Wilson index for a 19 year time span. First, we found that the activity level of  HD\,4915 was raising after the broad minimum reported in \cite{Shah_2018}. Second, we estimated the the calcium fluxes during this particular phase and found that exceed remarkably the basal level. Then, we computed the GLS periodogram and detected an activity cycle of 4.8-yr potentially modulated by a longer one. 

 All these facts provide evidence that the particular inactive phase detected in HD\,4915 could coincide with a minimum of the longer activity cycle, which suppresses the level of activity rather than indicates a MGM phase. Nevertheless, similar to \cite{Shah_2018}, we strongly suggest to continue observing HD\,4915 to confirm the longer cycle.
 
 To date, it is not an easy task to identify MGM candidates. We have some criteria  such as the relative variation of the $\mathrm{S}_\mathrm{{MW}}$ index around its mean ($\sigma_\mathrm{{S}}$/$\overline{\mathrm{S}}_\mathrm{{MW}}$), the decrease in the amplitude of consecutive cycles, and even the use of binary systems composed by similar stars \citep{Shah_2018,2021A&A...645L...6F,2022ApJ...936L..23L}. However, The main limitation remains the lack of extended time series. Consequently,  more activity surveys with extensive
records and suitable cadence are crucial for both the accurate identification of stars in Magnetic Grand Minima and the improvement of the current theoretical solar-dynamo models, including a better understanding Sun magnetic field.

\begin{acknowledgements}
PM, PC and JA acknowledge the financial support from the Consejo Nacional de Investigaciones Científicas y Técnicas (CONICET) in the form of doctoral fellowships. MJA thanks the financial support of the Dirección de Investigación y Desarrollo de la Universidad de La Serena (DIDULS/ULS, project ID: PAAI2021). CG
thanks the financial support of the Secretaría de Ciencia y Técnica in the form of CICITCA fellowships. MFT acknowledges funding from CONICET via Proyectos de Investigación bianual para investigadores asistentes y adjuntos (PIBAA, project ID: 28720210100242CO). AB thanks the Agencia I+D+i and the Universidad de Buenos Aires for their financial support. Finally, 
we thank the anonymous referee for their constructive comments, which allowed us to greatly improve the
quality of the manuscript.

\end{acknowledgements}

%
%

\bibliographystyle{aa}
\small
\bibliography{references}

\end{document}